\documentclass{moriond}
\usepackage{orcidlink}
\usepackage{float}

\def\be{\begin{equation}}
\def\ee{\end{equation}}
\def\bea{\begin{eqnarray}}
\def\eea{\end{eqnarray}}

\newcommand{\Photo}{\includegraphics[height=35mm, angle=-90]{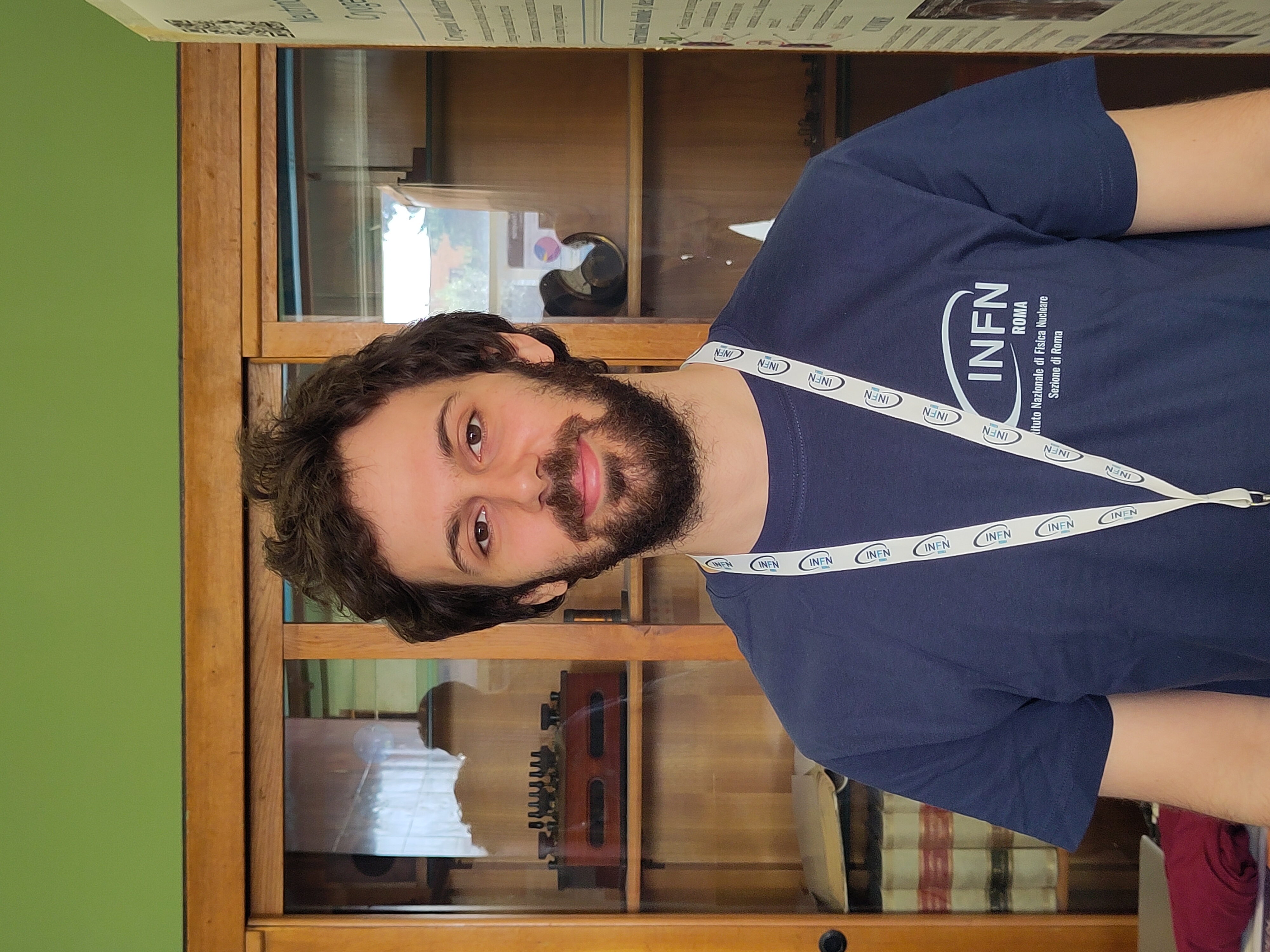}}

\begin{document}
\vspace*{4cm}
\title{A directed continuous-wave search from neutron stars in binary systems with the five-vector resampling technique}

\author{Francesco Amicucci\orcidlink{0009-0005-2139-4197}$^{1,2}$, Paola Leaci\orcidlink{0000-0002-3997-5046}$^{1,2}$, Pia Astone\orcidlink{0000-0003-4981-4120}$^{1}$, Sabrina D'Antonio\orcidlink{0000-0003-0898-6030}$^{^1}$, Stefano Dal Pra$^3$\orcidlink{0000-0002-1057-2307}, Matteo Di Giovanni$^{1,2}$\orcidlink{0000-0003-4049-8336}, Luca D'Onofrio\orcidlink{0000-0001-9546-5959}$^{1}$, Federico Muciaccia$^{1,2}$\orcidlink{0000-0003-0850-2649}, Cristiano Palomba\orcidlink{0000-0002-4450-9883}$^1$, Lorenzo Pierini$^{1}$\orcidlink{0000-0003-0945-2196}, Akshat Singhal$^{4}$\orcidlink{0000-0003-1275-1904}}
\address{$^1$INFN, Sezione di Roma, I-00185 Roma, Italy
	\\$^2$Università di Roma “La Sapienza", I-00185 Roma, Italy
	\\$^3$INFN, CNAF, 40127 Bologna, Italy
	\\$^4$HBCSE, Tata Institute of Fundamental Research, Mumbai 400088, India}
	
\maketitle\abstracts{
Continuous gravitational-wave signals (CWs), which are typically emitted by rapidly rotating neutron stars with non-axisymmetric deformations, represent particularly intriguing targets for the Advanced LIGO-Virgo-KAGRA detectors. These detectors operate within sensitivity bands that encompass more than half of the known pulsars in our galaxy existing in binary systems, which are the targeted sources of this paper. However, the detection of these faint signals is especially challenged by the Doppler modulation due to the source's orbital motion, typically described by five Keplerian parameters, which must be determined with high precision to effectively detect the signal. This modulation spreads the signal across multiple frequency bins, resulting in a notable reduction of signal-to-noise ratio and potentially hindering signal detection. To overcome this issue, a robust five-vector resampling data-analysis algorithm has been developed to conduct thorough directed/narrowband CW searches at an affordable computational cost. We employ this methodology for the first time to search for CWs from Scorpius X-1, using publicly available data from the third observing run of the Advanced LIGO-Virgo-KAGRA detectors. No statistically significant CW signals can be claimed. Hence, we proceeded setting 95\% confidence-level upper limits in selected frequency bands and orbital parameter ranges, while also evaluating overall sensitivity.}

\section{Introduction}

Unlike transient Gravitational Waves (GWs) emitted by the merger of compact objects, there are much fainter, persistent/continuous GWs (CWs), which are being continuously emitted by a source that has quasi-periodic rotational frequency. Due to their weakness, we need to integrate for long observing times to accumulate an adequate signal-to-noise-ratio~\cite{universe5110217}. Promising sources of CWs include neutron stars (NSs), the remnants of massive stellar cores after a supernova explosion, that can emit a continuous stream of GWs due to their asymmetrical shapes, non-uniform rotation, or other structural irregularities~\cite{PhysRevD.66.084025}. The most commonly accepted model suggests \textit{mountains} supported by elastic and/or magnetic stresses, where the NS rotational frequency $f_{\rm rot}$ is related to the GW frequency $f_{\rm GW}$ by $f_{\rm GW}=2f_{\rm rot}$ \cite{PhysRevD.20.351}. Detecting CWs would allow us to get a set of information that electromagnetic observations alone cannot achieve, such as information on NS quadrupolar deformation (i.e., ellipticity), NS properties (i.e., the range of NS masses, radii, population models), and in general cold dense matter Equation Of State properties \cite{universe5110217}.

Scorpius X-1 is the brightest Low-Mass X-ray binary (LMXB) composed of a NS in a binary orbit with a low-mass normal star (i.e., its companion). A characteristic property of these systems is the inflow of gas from the companion star to the NS, in a process known as accretion \cite{10.1093/mnras/184.3.501,Bildsten_1998}. The by-product of accretion is the generation of X-rays: the more accretion, the more X-rays are produced.

An interesting astrophysical model predicts that the GW strain $h_0$ from a LMXB at the torque-balance level is proportional to the square root of the x-ray flux~\cite{1984ApJ_278_345W,Bildsten_1998,10.1093/mnras/184.3.501}. Since Scorpius X-1 is the brightest LMXB known so far, we consider it as one of the most interesting potential CW sources. On the other hand, the NS frequency rotation is also unknown, and so is also $f_{\rm GW}$. As a result, searches targeting Scorpius X-1 must cover a wide range of the interferometer power spectrum, which significantly increases the associated computational cost.

\section{The signal and search setup}
As discussed in~\cite{Singhal_2019}, a monochromatic GW signal undergoes multiple time-dependent modulations when it reaches the ground-based detector. Our goal is to compute the time series in a new time coordinate $t'$, using the \textit{stroboscopic resampling}, i.e., the time series is downsampled at irregular intervals where the phase exhibits linear behavior. In other words, assuming that the original time series is evenly spaced, it is downsampled at the nearest integer values of the new time variable $t'$. It is important to note that, in the absence of spindown terms, stroboscopic resampling can be performed without prior knowledge of the frequency, as the time correction is independent of it. 

The (monochromatic) signal phase in the (NS) source frame is
\begin{eqnarray}
	\Phi^{\rm NS}(\tau)=\phi_0+2\pi f_{\rm GW}(\tau-\tau_0)\,,
\end{eqnarray}

where $\tau$ is the source emission time and $\phi_0$ is the phase at the reference time $\tau_0$.

The signal phase observed at the detector frame is
\begin{eqnarray}\label{eq:phase-det}
	\Phi^{\rm det}(t_{\rm arr})=\phi_0+2\pi f_{\rm GW}\left(t_{\rm arr}+\Delta\tau(t_{\rm arr})-\tau_0\right),
\end{eqnarray}

in $\Delta\tau$ are parametrized all the time delays that occurrs due to
the motion of the detector with respect to the source (Doppler effect). In particular, the binary Rømer delay~\cite{PhysRevD.91.102003,Leaci_2017} $R(t)$ can be expressed in terms of five Keplerian parameters $\{P,a_{\rm p}, e, \omega, t_{\rm p}\}$, where $a_{\rm p}$ is the projected semi-major axis in ls, $P$ is the time period required by a NS to complete its orbit around the Binary System Barycentre, $e$ is the eccentricity, $\omega$ is the argument of periapsis, and $t_{\rm p}$ is the time of periapsis passage. In the case of low eccentric orbit ($e\ll 1$), $R(t)$ can be equivalently parametrized by $\{\Omega, a_{\rm p}, \kappa, \eta, t_{\rm asc}\}$, where $\eta\equiv e\sin\omega$ and $\kappa\equiv e\cos\omega$ represent the \textit{Laplace-Lagrange} parameters, $t_{\rm asc}$ is the time of ascending node and $\Omega$ is the mean orbital angular velocity. Other smaller effects are also considered in the signal model, namely the Einstein and the Shapiro delays. Hence, we can introduce the new time variable $t'=t_{\rm arr}+\Delta\tau(t_{\rm arr})$  such that the phase $\Phi^{\rm det}(t')$ is equal to the signal with constant frequency, i.e., we can remove all modulations. This means that a single correction applies to all frequencies, making it especially useful for searching for a signal from Scorpius X-1, where the spin frequency is unknown.

The amplitude of the signal measured at the detector has also a modulation due to the sidereal motion, producing a splitting of the frequency into five harmonics at  $f_{\rm GW},f_{\rm GW}\pm f_{\rm sd},f_{\rm GW}\pm 2f_{\rm sd}$, where $f_{\rm sd}$ is the sidereal frequency\footnote{The Earth sidereal frequency is the inverse of sidereal period, which is $\approx86164$ s ($\approx 23 {\rm  h }\; 56 {\rm m }\; 4 {\rm s}$).} ($1.16 \times 10^{-5}$ Hz). The acccuracy on the $\{P,a_{\rm p}, e, \omega, t_{\rm p}\}$ parameters determines the ability to recover the 5 peaks.

\section{Results}
The uncertainties in the electromagnetically measured binary parameters of Scorpius X-1 are significantly larger than the maximum offsets in these parameters that we can use to detect a true CW signal without deliberately resorting to a matched-filtering approach, as indicated by a previous study \cite{Singhal_2019}. Hence, we restrict our search to three orbital parameter configurations: one assuming the Scorpius X-1 ephemerides $\{P,a_p, e, \omega, t_p\}$ are exact, and two additional configurations where the parameters are adjusted by their respective maximum offsets, i.e., $\{P\pm \Delta P,a_p\pm \Delta a_p,e \pm \Delta e,\omega\pm \Delta \omega,t_p \pm \Delta t_p\}$, with $\{\Delta P = 24\,{\rm ms}, \Delta a_p = 14\,{\rm mls}, \Delta e = 10^{-4}, \Delta \omega = 0.01^\circ, \Delta t_p = 0.1\,{\rm s}\}$ being the maximum values of binary-parameter offsets. These searches spanned the entire frequency range of $[10-1000]$ Hz and the full O3 dataset\footnote{O3 started on April 1, 2019 15:00:00 UTC (GPS 1238166018) and ended on March 27, 2020 17:00:00 UTC (GPS 1269363618).}. Further details can be found here~\cite{amicucci2025directedcontinuouswavesearchscorpius}. Since none of the selected candidates survived the latest stages of the follow-up and veto steps, no detection of CW signals can be claimed. However, upper limit curves on the signal amplitude $h_0$ can still be derived. These curves define a threshold above which the presence of a CW signal can be excluded with a certain confidence level. The most stringent upper limit value is achieved at  $f \approx 229.5$ Hz and is equal to $h_0 \approx 5.64\times 10^{-26}$ for the Livinston data.

\begin{figure}[H]

		\includegraphics[width=0.9\linewidth]{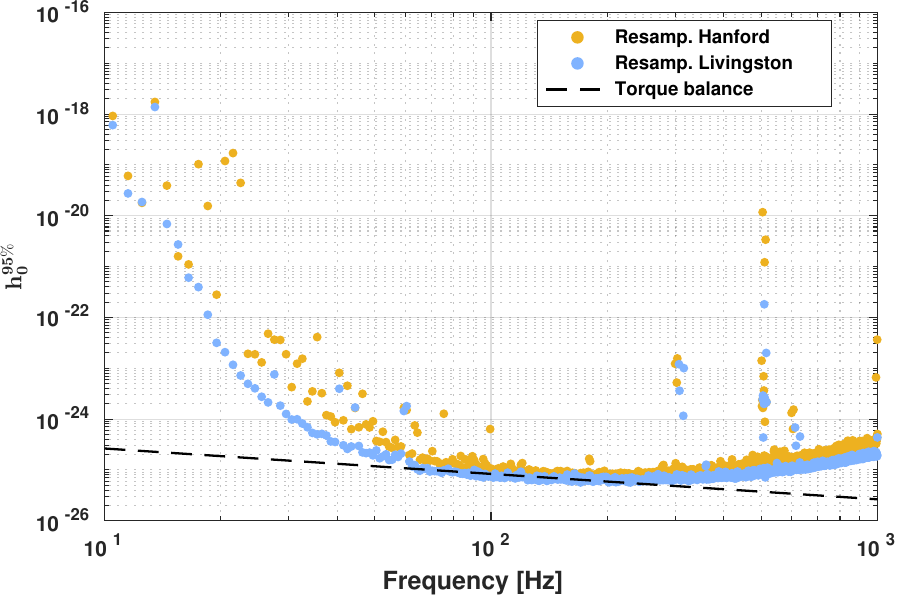}

	\caption[Sensitivity estimation]{Sensitivity estimation as a function of frequency for the Hanford (orange points) and Livingston (blue points) detectors. These estimates are extrapolated from the computed upper limits with the maximum detection statistic values found in every analyzed 1 Hz-wide band, see~\cite{amicucci2025directedcontinuouswavesearchscorpius} for further details. The Upper limits are computed in selected frequency bands and orbital parameter ranges. We note that the theoretical strain at the torque-balance level (black dashed line) is obtained assuming $M_{{\rm NS}}=1.4 M_{\odot}$ and $R_{{\rm NS}}\approx 10\, { \rm km}$.
	}
	\label{fig:Sens}
\end{figure}

As discussed in~\cite{Abbott_2022_scox1}, deriving firm conclusions about the NS equation of state or magnetic field strength from our upper limits is challenging without assuming a specific accretion model. The resulting sensitivity estimation versus frequency results are shown in Figure~\ref{fig:Sens}, where it is also shown the theoretical strain (at the torque-balance level) assuming $M_{{\rm NS}}=1.4 M_{\odot}$ and $R_{{\rm NS}}\approx 10 { \rm km}$ \cite{Singhal_2019}
\begin{eqnarray}\label{eq:expc-strain}
	h_0 \approx 3.5 \times 10^{-26}\sqrt{\frac{300 {\rm Hz}}{f_{{\rm rot}}}}\,.
\end{eqnarray}
The ``torque balance" curve is included solely to facilitate comparison with existing literature.

\section*{Acknowledgments}

This material is based upon work supported by NSF's LIGO Laboratory which is a major facility fully funded by the National Science Foundation. The authors are grateful for computational resources provided by the CNAF and supported by INFN.

\section*{References}
\bibliography{moriond}

%%% manually generated bibliography
%\begin{thebibliography}{99}
%\bibitem{ja}C Jarlskog in {\em CP Violation}, ed. C Jarlskog
%(World Scientific, Singapore, 1988).
%\bibitem{ma}L. Maiani, \Journal{\PLB}{62}{183}{1976}.
%\bibitem{bu}J.D. Bjorken and I. Dunietz, \Journal{\PRD}{36}{2109}{1987}.
%\bibitem{bd}C.D. Buchanan {\it et al}, \Journal{\PRD}{45}{4088}{1992}.
%\end{thebibliography}

\end{document}